\documentclass[]{spie}  

\usepackage{amsmath,amsfonts,amssymb}
\usepackage{graphicx}
\usepackage{multirow}
\usepackage{adjustbox}
\usepackage[colorlinks=true, allcolors=blue]{hyperref}
\usepackage{caption} 
\title{Enhancing Hierarchical Transformers for Whole Brain Segmentation with Intracranial Measurements Integration}

\author[a]{Xin Yu}
\author[b,c]{Yucheng Tang}
\author[a]{Qi Yang}
\author[a]{Ho Hin Lee}
\author[b]{Shunxing Bao}
\author[a,b]{Yuankai Huo}
\author[a,b]{Bennett A. Landman}
\affil[a]{Computer Science, Vanderbilt University, Nashville, TN, USA}
\affil[b]{Electrical and Computer Engineering,  Vanderbilt University, Nashville, TN, USA}
\affil[c]{Nvidia Corporation}

\begin{document} 
\maketitle

\begin{abstract}
Whole brain segmentation with magnetic resonance imaging (MRI) enables the non-invasive measurement of brain regions, including total intracranial volume (TICV) and posterior fossa volume (PFV). Enhancing the existing whole brain segmentation methodology to incorporate intracranial measurements offers a heightened level of comprehensiveness in the analysis of brain structures. Despite its potential, the task of generalizing deep learning techniques for intracranial measurements faces data availability constraints due to limited manually annotated atlases encompassing whole brain and TICV/PFV labels. In this paper, we enhancing the hierarchical transformer UNesT for whole brain segmentation to achieve segmenting whole brain with 133 classes and TICV/PFV simultaneously. To address the problem of data scarcity, the model is first pretrained on 4859 T1-weighted (T1w) 3D volumes sourced from 8 different sites. These volumes are processed through a multi-atlas segmentation pipeline for label generation, while TICV/PFV labels are unavailable. Subsequently, the model is finetuned with 45 T1w 3D volumes from Open Access Series Imaging Studies (OASIS) where both 133 whole brain classes and TICV/PFV labels are available. We evaluate our method with Dice similarity coefficients(DSC). We show that our model is able to conduct precise TICV/PFV estimation while maintaining the 132 brain regions performance at a comparable level. Code, trained model, and Singularity are available at: \url{https://github.com/MASILab/UNesT/tree/main/wholebrainSeg}.

\end{abstract}

\keywords{Intracranial measurements, Whole brain segmentation, Hierarchical transformer}

\section{INTRODUCTION}
\label{sec:intro}  

Whole brain segmentation through magnetic resonance imaging (MRI) provides the means for non-invasively measuring various brain regions and supports clinical investigations aimed at enhancing our understanding of the complexities of the human brain \cite{huo20193d}. Among the brain regions,  total intracranial volume (TICV) refer to the volume contained within the skull, encompassing the brain, meninges, and cerebrospinal fluid (CSF). It is an vital covariate in the volumetric analysis of brain and brain region in particular the investigation of neurodegenerative disorders~\cite{malone2015accurate,liu2022generalizing}. Posterior fossa volume (PFV) has the potential to serve as an indicator for early diagnosis of Chiari type I malformation (CMI) \cite{vurdem2012analysis,atkinson1998evidence,badie1995posterior}. Hence, it is important to encompass TICV and PFV within the whole brain segmentation pipeline. 

Traditionally, researchers utilized atlases for brain segmentation~\cite{hansen2021pandora,yang2022learning}, leveraging anatomical priors. With the advent of deep learning in medical imaging~\cite{ronneberger2015u,hatamizadeh2022unetr,yang2022quantification,yu2023longitudinal,yu2020deep,yang2022label,yu2022reducing,yang2023single,lee2023scaling,yu2022accelerating}, the focus has been shifted, replacing manual features with automatically learned features by computers. This transition has demonstrated impressive performance in brain segmentation tasks, prompting extensive efforts in this direction.\cite{de2015deep,wachinger2018deepnat,dey2018compnet,hansen2021pandora,yang2022learning}. In particular, Huo \textit{et al.} \cite{huo20193d} proposed a tile-based approach, called SLANT, for segmenting 132 brain regions. They partitioned the entire brain into 27 tiles, each processed by distinct U-Nets \cite{cciccek20163d}, and aggregated the ensemble outcomes of these 27 U-Nets to produce the final results. Liu \textit{et al.} \cite{liu2022generalizing} further improve the SLANT to incorporate TICV/PFV segmentation by adding two additional channels in the final output layer. Recent years, vision transformer-based models \cite{liu2021swin,dosovitskiy2020image,tang2022self} have exhibited remarkable representation learning capabilities in computer vision and medical image analysis by effectively capturing global dependencies. 
However, as transformer-based approaches reshape images into 1D sequences, maintaining positional information among patches becomes challenging. 
Yu \textit{et al.} \cite{yu2023unest} proposed a 3D hierarchical transformer-based method, named UNesT, that allows local patch communication. This method effectively models 132 brain regions using a single model, contrasting with SLANT's 27 models, and attains the state-of-the-art performance in whole brain segmentation. However, this method lacks TICV/PFV estimation, which restricts its applicability for downstream analysis.

In this study, we enhance the UNesT framework by incorporating intracranial measurements. Specifically, we integrate TICV/PFV estimation by introducing two additional convolutional layers. These layers simultaneously process the TICV and PFV outputs alongside the other 132 brain regions. To address the data scarcity problem, we follow the practice in UNesT that pretraining the model with a large dataset with pseudo labels and finetune with human annotations. We use 45 T1w 3D volumes from Open Access Series Imaging Studies (OASIS)~\cite{marcus2007open} to finetune the model, where both 133 whole brain classes and TICV/PFV labels are available. Our results show that we can achieve accurate TICV/PFV estimation while maintaining a comparable level of performance across 132 regions on whole brain segmentation.


\section{Methods}
\begin{figure*}[h!]
  \centering
\includegraphics[width=1\textwidth]{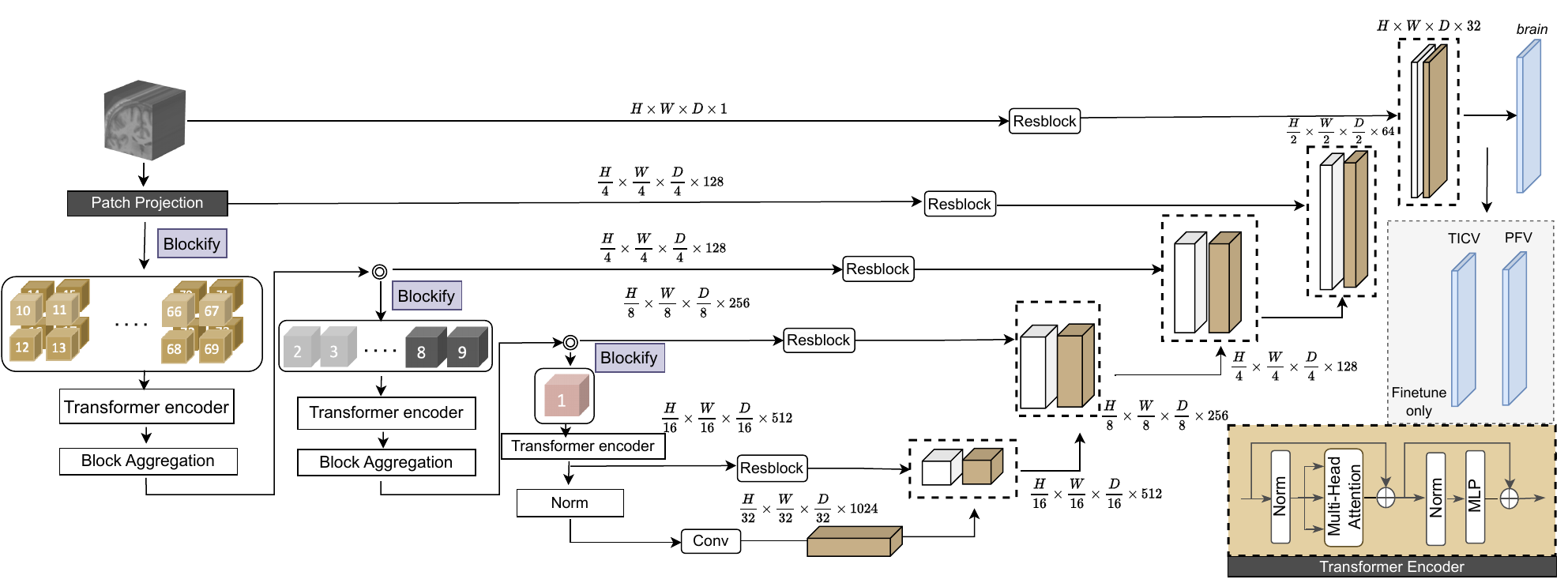}
\caption{Overview of the model architecture. Input image patch sequence are aggregated into block in each hierarchy and fed into transformer block separately. Blocks are deblockified back to image patch sequence space at the end of each hierarchy for inter-blocks communication. In the pretraining stage, the final feature maps is transformed to have 133 channels. In the finetuning stage, the final feature maps is transformed to 3 different feature maps, each with 133, 1, and 1 channels respectively, to facilitate segmentation for 132 brain regions, TICV, and PFV.}
\label{fig:fig1}
\end{figure*}

The approach consists of two-stage: pretraining using pseudo labels for 132 brain regions, followed by refinement through finetuning involving brain region as well as TICV/PFV labels, as shown in Figure~\ref{fig:fig1}.

\subsection{Data}
4859 T1w MRI from eight different sites have been incorporated for pretraining (Table~\ref{tab:dataset}). The labels for these scans have been generated using an established multi-atlas segmentation pipeline~\cite{asman2014hierarchical}, while TICV/PFV labels are unavailable. 

\begin{table*}[ht]
\centering
\caption{Summary of the data from the 4859 multi-site images~\cite{yu2023unest}.}
\label{tab:dataset}
\begin{tabular}{lll}
\hline
Site Name                                   & Number \\ \hline
Baltimore Longitudinal Study of Aging (BLSA)         & 614    \\
Cutting Pediatrics                                   & 586    \\
Autism Brain Imaging Data Exchange (ABIDE)          & 563    \\
Information Extraction from Images (IXI)             & 541    \\
Attention Deficit Hyperactivity Disorder (ADHD200)  & 950    \\
Open Access Series on Imaging Study (OASIS)          & 312    \\
1000 Functional Connectome (fcon\_1000)               & 1102   \\
Nathan Kline Institute Rockland (NKI\_rockland)    & 141    \\ \hline
\end{tabular}
\end{table*}

 45 T1w MRI volumes from OASIS are used to finetune the model. This dataset provides both 133 whole brain classes and TICV/PFV labels. The 133 brain labels are manually traced based on the BrainCOLOR protocol. All T1w volumes are registered to the MNI space using the MNI305 template \cite{evans19933d}. The registered volume have size of 172 $\times$ 220 $\times$ 156 with isotropic spacing of 1mm.

\subsection{UNesT}
The input volumes, $\mathcal{X} \in \mathbb{R}^{H\times{W}\times{D}\times{C}}$, are tokenized into size of $\frac{H}{P_h} \times \frac{W}{P_w} \times \frac{D}{P_d} \times E$ where $P_h \times P_w \times P_d $ represent patch size and $E$ represent the embedded size. To achieve local patch communication, the adjacent patches sequence are aggregated together to form a block. The input volume is reshaped to $\mathcal{X} \in \mathbb{R}^{b \times T \times{n} \times C^{'}}$, where $T$ is the number of blocks, $b$ is the batch size, and $n$ is the total length of sequences. Each block is fed into the transformer block consisting layer normalization (LN), multi-head self-attention (MSA), and multi-layer perceptron (MLP) separately so that the communication among patches are performed locally within each block. The blocks are aggregated and transformed back to the original patch sequences space after the transformer block. The communication between different blocks is performed during the aggregation where the volume size are reduced by a factor of 2 in each direction. This marks the initial level of the transformer encoder hierarchy. This repetition continues until the count of blocks reduces to 1, resulting in the formation of three hierarchical levels. 

The decoder started with the output of the last hierarchy and upsamples the feature map as in 3D U-Net \cite{cciccek20163d}. Specifically, the outcome feature map from each hierarchical stage, at various resolutions, is incorporated into the decoder's feature map using a skip connection. 

\subsection{TICV/PFV Estimation}
In the pretraining stage, the final convolutional layer of the network transforms the feature map with 32 channels to 133 channels and uses softmax to obtain the segmentation mask. In the finetuning stage, two extra convolutional layers are introduced. These layers have the purpose of transforming the 32-channel feature map into two distinct feature maps, each having a single channel. The segmentation masks for TICV and PFV are derived by applying the sigmoid function to these feature maps. Throughout the fine-tuning process, optimization is performed on both the segmentation mask for the 133 brain classes and the TICV/PFV segmentation masks. Dice loss is used to optimize both the 132 brain regions and Dice and binary cross entropy loss are used to optimize the TICV/PFV:
\begin{equation}
L= L_{brain} + \beta_1 L_{TICV} + \beta_2 L_{PFV},
\label{equation}
\end{equation}
where $L_{brain}$, $L_{TICV}$ and $L_{PFV}$ represent the Dice loss for 133 brain classes, TICV and PFV, respectively. $\beta_1$ and $\beta_2$ are the weights for the loss for TICV and PFV, respectively.

\section{Experiments and Results}

\subsection{Implementation Details}
The experiments are conducted using PyTorch and MONAI frameworks. Models are trained with single Nvidia RTX 5000 16G GPU. The input volume is randomly cropped to 96 $\times$ 96 $\times$ 96 during online augmentation. No additional augmentation methods are employed, as our empirical results shows that including augmentation leads to a decrease in model performance. For pre-training with auxiliary labels, the initial learning rate is 0.0001, with a weight decay of $1e^{-5}$, and the training is conducted over 200K iterations. We partitioned the set of 45 T1w images with TICV/PFV labels into subsets of 32, 8, and 5 for training, validation, and held-out testing, respectively. During the finetuning phase, the learning rate is adjusted to $1e^{-5}$ for a training duration of 25K iterations. The learning rate is adjusted by a cosine scheduler. In the first 20K iteration of the finetuning step, in $\beta_1$ and $\beta_2$ in Equation~\ref{equation} are set to be 0.8 and 1.0, respectively. For the iteration after 20K, $\beta_1$ and $\beta_2$ are reduced to 0.08 and 0.1, respectively.
\subsection{Experimental and Discussion}
\begin{figure*}[h!]
  \centering
  \includegraphics[width=0.6\textwidth]{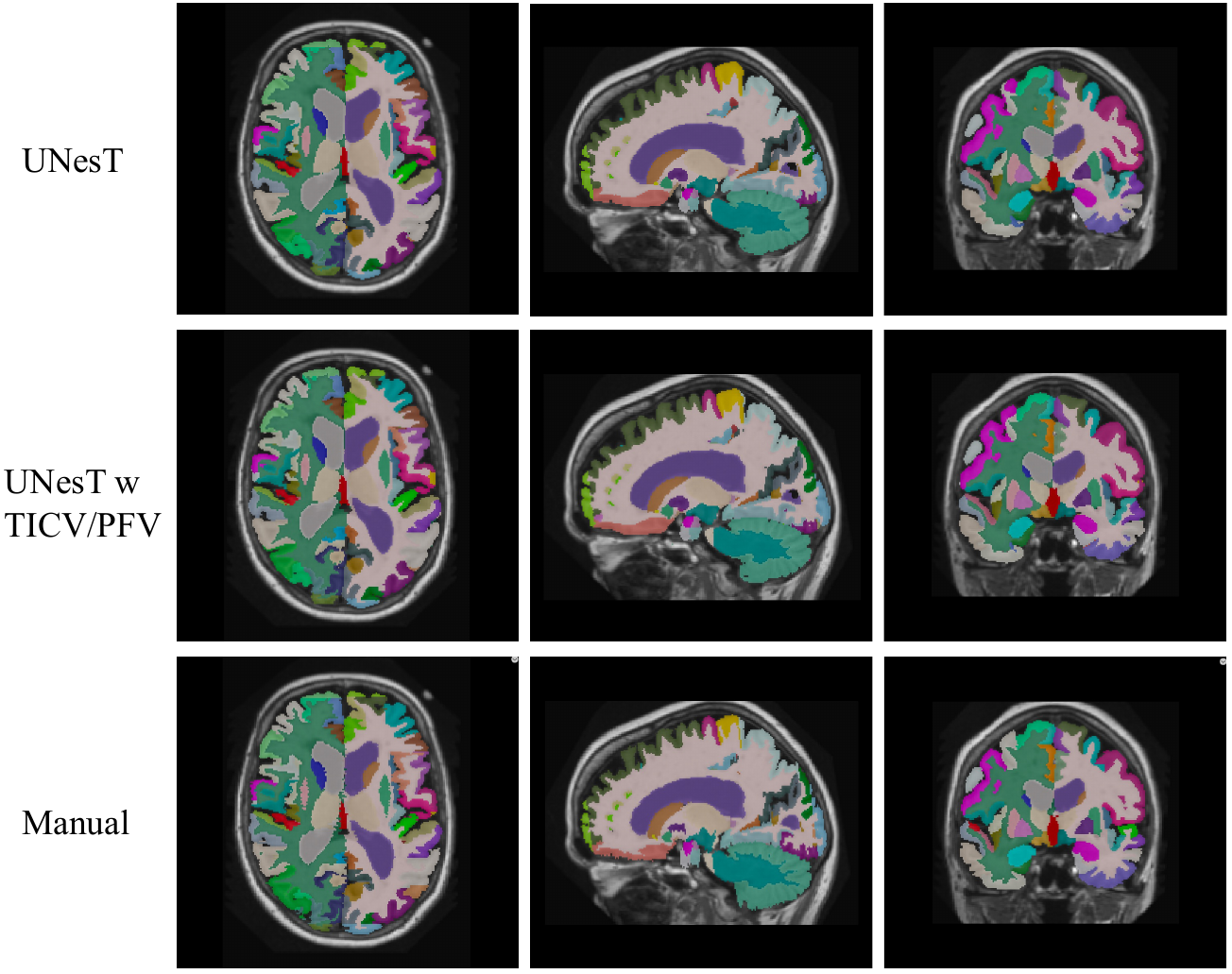}
\caption{Visualization of the 132 brain regions segmentation on a randomly selected case in the test set.}
\label{fig:fig3}
\end{figure*}

\begin{figure*}[h!]
  \centering
  \includegraphics[width=\textwidth]{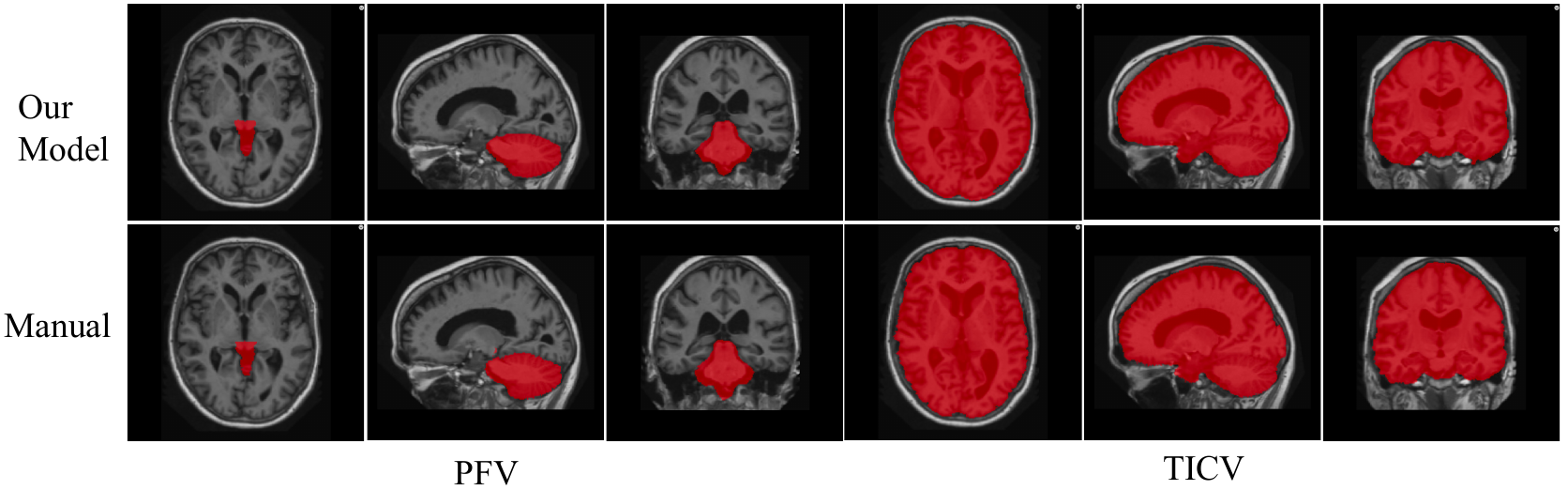}
\caption{Visualization of the TICV/PFV segmentation on a randomly selected case in the test set.}
\label{fig:fig4}
\end{figure*}

We evaulate the model performance with Dice similarity coefficients(DSC). The segmentation results of 132 brain regions and TICV/PFV from the plain UNesT and our model UNest extenstion is shown in Table~\ref{results}, Figure~\ref{fig:fig3} and Figure~\ref{fig:fig4}. We show that we can achieve accurate TICV/PFV segmentation, reflected in DSC scores of 0.962 and 0.954. When comparing the segmentation performance for the 132 brain regions, our achieved DSC score of 0.751 closely aligns with the plain UNesT's performance level of 0.759.

\begin{table}[]
\centering
\caption{DSC score for UNesT and UNesT with TICV/PFV estimation on 132 brain regions, TICV and PFV. Note: LCI and UCI represents the lower and upper bounds of the 95\% confidence interval, respectively.}
\label{results}
\begin{tabular}{c|cll|ccl|cll}
\hline
\multirow{2}{*}{Model} & \multicolumn{3}{c|}{Brain}              & \multicolumn{3}{c|}{TICV}                                            & \multicolumn{3}{c}{PFV}                                       \\ \cline{2-10} 
                       & \multicolumn{1}{l}{DSC} & LCI   & UCI   & \multicolumn{1}{l}{DSC} & \multicolumn{1}{l}{LCI} & UCI              & \multicolumn{1}{l}{DSC} & LCI              & UCI              \\ \hline
UNesT                  & 0.759                   & 0.735 & 0.784 & \textbackslash{}        & \textbackslash{}        & \textbackslash{} & \textbackslash{}        & \textbackslash{} & \textbackslash{} \\
UNesT w TICV/PFV       & 0.751                   & 0.728 & 0.773 & 0.962                   & 0.956                   & 0.968            & 0.954                   & 0.946            & 0.962            \\ \hline
\end{tabular}
\end{table}

Our experiments illustrate that our model can effectively achieve precise TICV/PFV segmentation while maintaining performance in the segmentation of the 132 brain regions. However, it is worth noting that during training, the DSC score for TICV validation reaches approximately 0.93 at around 2K iterations, while PFV reaches about 0.93 at around 10K iterations. The model prioritizes optimizing TICV/PFV, leading to a notable decrease in the performance of the 132 brain regions segmentation comparing with model trained without TICV/PFV estimation. Hence, we reduce the weight of $\beta_1$ and $\beta_2$ after 20K iterations.

\begin{figure*}[h!]
  \centering
  \includegraphics[width=0.7\textwidth]{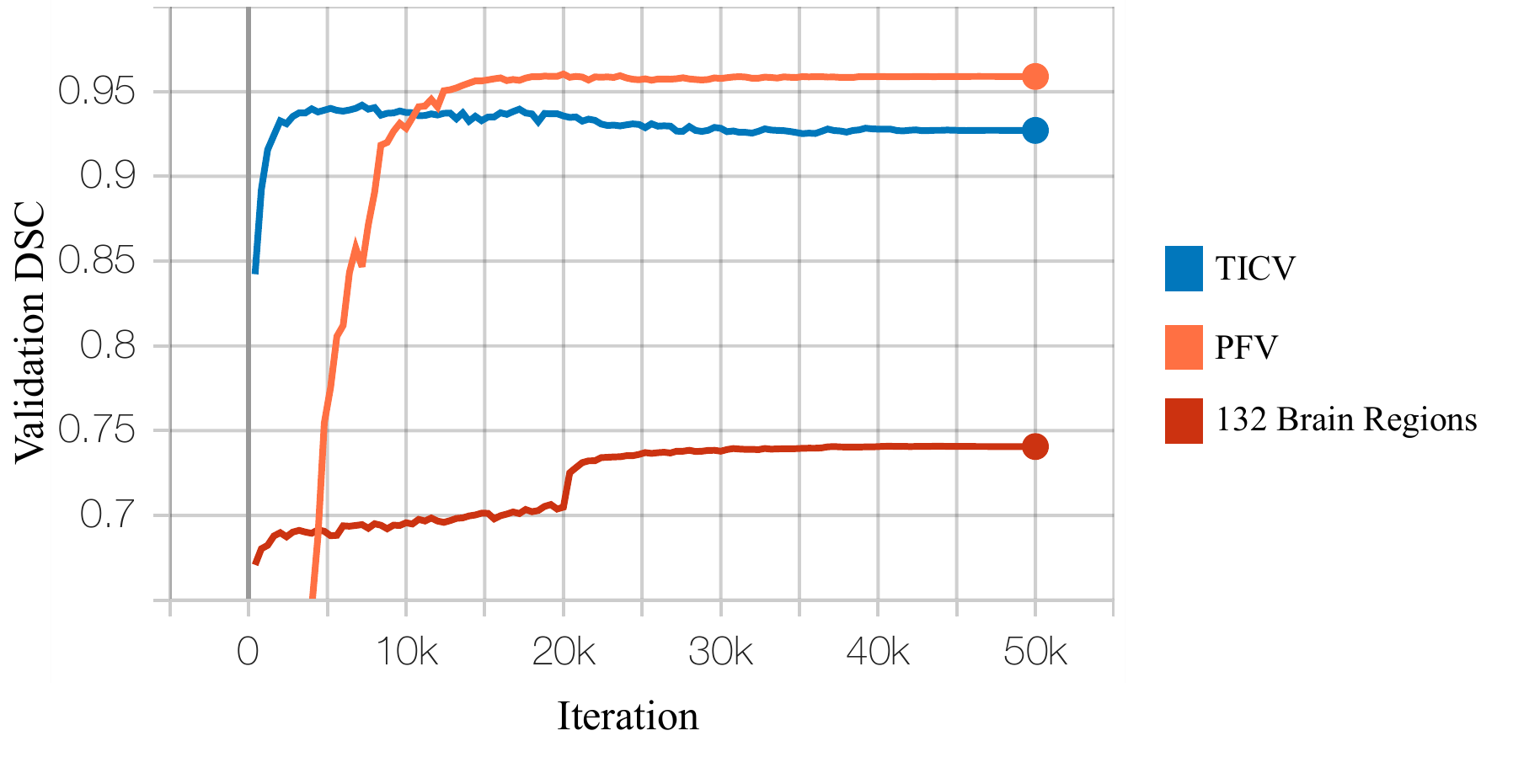}
\caption{Training curve on the validation set. After reducing the weight of TICV/PFV at 20K iteration, the performance of TICV drops gradually. However, the performance of the 132 brain regions shows a marked improvement.}
\label{fig:fig2}
\end{figure*}

In Figure~\ref{fig:fig2}, we can see that the performance of TICV drops gradually after reducing the weight of TICV/PFV at 20k iterations. However, the performance of the 132 brain regions shows a marked improvement. This observation suggests that TICV does not complement the segmentation of the 132 brain regions. This could be due to a fundamental conflict in objectives between segmenting the 132 brain regions and the TICV task. The primary goal of segmenting the 132 brain regions is to achieve maximal separation between each region, whereas the TICV task requires combining several regions together.

\section{Containerized Implementation}
\begin{figure*}[h!]
  \centering
\includegraphics[width=\textwidth]{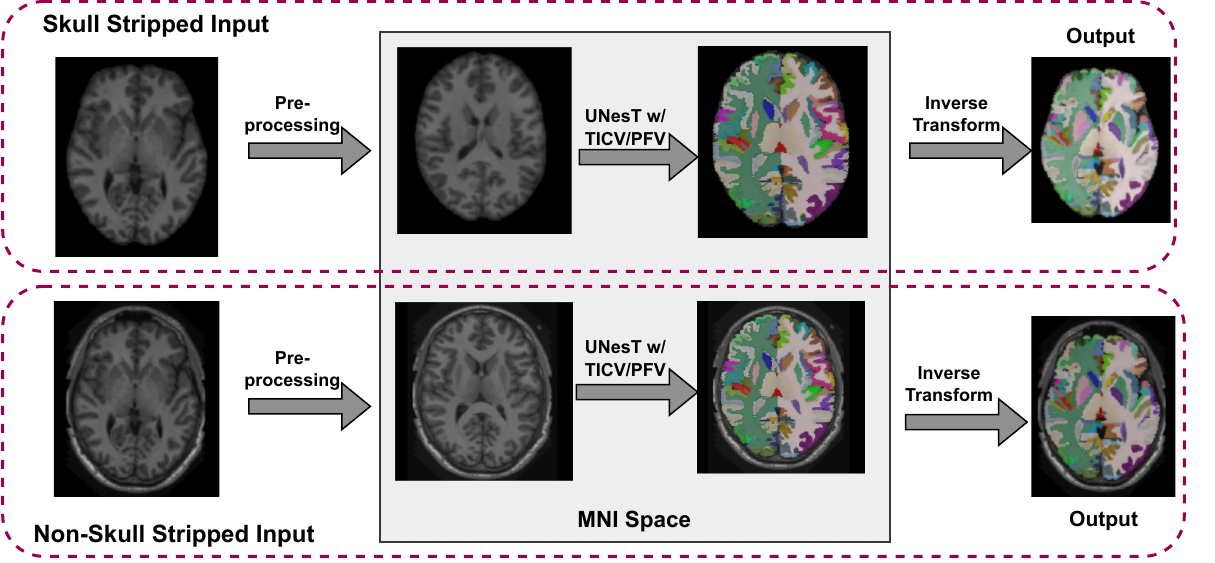}
\caption{Overview of the Singularity workflow. The Singularity can accommodate both skull-stripped input and non-skull stripped input.}
\label{fig:fig5}
\end{figure*}

For the ease of usage, we have built a containerized solution using Singularity for end-to-end segmentation with single command. The download link and instructions can be found at: \url{https://github.com/MASILab/UNesT/tree/main/wholebrainSeg}. Our Singularity can accommodate both skull striped and non-skull striped data. The models for skull-stripped data are trained using data that has undergone skull stripping with the synthstrip tool provided in Freesurfer \cite{segonne2004hybrid}. The workflow of the Singularity is shown in Figure~\ref{fig:fig5}. The input data undergoes pre-processing step including N4 bias field correction, intensity normalization and affine registration to MNI space~\cite{huo20193d}. The enhanced UNesT model is then applied to the processed data, producing a segmentation map within the MNI space. Subsequently, an inverse transform is applied to bring the segmentation map back to the original space, resulting in the final segmentation output.

\section{Conclusion}
Herein, we enhance the current hierarchical transformer UNesT for whole brain segmentation by integrating intracranial measurements. More precisely, we include TICV/PFV estimation by introducing an extra set of convolutional layers. These additional layers enable the estimation of TICV and PFV segmentation masks alongside the other 132 brain regions. Importantly, this is achieved while ensuring that the performance of these 132 brain regions remains at a comparable level. This work expands the potential usage of UNesT to various other downstream analyses.

\section*{Acknowledgments}
This research is supported by NIH Common Fund and National Institute of Diabetes, Digestive and Kidney Diseases U54DK120058, NSF CAREER 1452485, NIH grants, 2R01EB006136, 
1R01EB017230 (Landman), R01NS09529, and R01DK135597 (Huo). The identified datasets used for the analysis described were obtained from the Research Derivative (RD), database of clinical and related data. The imaging dataset(s) used for the analysis described were obtained from ImageVU, a research repository of medical imaging data and image-related metadata. ImageVU and RD are supported by the VICTR CTSA award (ULTR000445 from NCATS/NIH) and Vanderbilt University Medical Center institutional funding. ImageVU pilot work was also funded by PCORI (contract CDRN-1306-04869).
\bibliography{report} 

\begin{thebibliography}{10}

\bibitem{huo20193d}
Huo, Y., Xu, Z., Xiong, Y., Aboud, K., Parvathaneni, P., Bao, S., Bermudez, C.,
  Resnick, S.~M., Cutting, L.~E., and Landman, B.~A., ``3d whole brain
  segmentation using spatially localized atlas network tiles,'' {\em
  NeuroImage}~{\bf 194},  105--119 (2019).

\bibitem{malone2015accurate}
Malone, I.~B., Leung, K.~K., Clegg, S., Barnes, J., Whitwell, J.~L., Ashburner,
  J., Fox, N.~C., and Ridgway, G.~R., ``Accurate automatic estimation of total
  intracranial volume: a nuisance variable with less nuisance,'' {\em
  Neuroimage}~{\bf 104},  366--372 (2015).

\bibitem{liu2022generalizing}
Liu, Y., Huo, Y., Dewey, B., Wei, Y., Lyu, I., and Landman, B.~A.,
  ``Generalizing deep learning brain segmentation for skull removal and
  intracranial measurements,'' {\em Magnetic resonance imaging}~{\bf 88},
  44--52 (2022).

\bibitem{vurdem2012analysis}
Vurdem, {\"U}.~E., Acer, N., Ertekin, T., Savranlar, A., Inci, M.~F., et~al.,
  ``Analysis of the volumes of the posterior cranial fossa, cerebellum, and
  herniated tonsils using the stereological methods in patients with chiari
  type i malformation,'' {\em The Scientific World Journal}~{\bf 2012} (2012).

\bibitem{atkinson1998evidence}
Atkinson, J., Kokmen, E., and Miller, G.~M., ``Evidence of posterior fossa
  hypoplasia in the familial variant of adult chiari i malformation: case
  report.,'' {\em Neurosurgery}~{\bf 42}(2),  401--3 (1998).

\bibitem{badie1995posterior}
Badie, B., Mendoza, D., and Batzdorf, U., ``Posterior fossa volume and response
  to suboccipital decompression in patients with chiari i malformation,'' {\em
  Neurosurgery}~{\bf 37}(2),  214--218 (1995).

\bibitem{hansen2021pandora}
Hansen, C.~B., Yang, Q., Lyu, I., Rheault, F., Kerley, C., Chandio, B.~Q.,
  Fadnavis, S., Williams, O., Shafer, A.~T., Resnick, S.~M., et~al., ``Pandora:
  4-d white matter bundle population-based atlases derived from diffusion mri
  fiber tractography,'' {\em Neuroinformatics}~{\bf 19},  447--460 (2021).

\bibitem{yang2022learning}
Yang, Q., Hansen, C.~B., Cai, L.~Y., Rheault, F., Lee, H.~H., Bao, S., Chandio,
  B.~Q., Williams, O., Resnick, S.~M., Garyfallidis, E., et~al., ``Learning
  white matter subject-specific segmentation from structural mri,'' {\em
  Medical physics}~{\bf 49}(4),  2502--2513 (2022).

\bibitem{ronneberger2015u}
Ronneberger, O., Fischer, P., and Brox, T., ``U-net: Convolutional networks for
  biomedical image segmentation,'' in [{\em Medical Image Computing and
  Computer-Assisted Intervention--MICCAI 2015: 18th International Conference,
  Munich, Germany, October 5-9, 2015, Proceedings, Part III
  18}{\nolinebreak\hspace{0.1em}]},   234--241, Springer (2015).

\bibitem{hatamizadeh2022unetr}
Hatamizadeh, A., Tang, Y., Nath, V., Yang, D., Myronenko, A., Landman, B.,
  Roth, H.~R., and Xu, D., ``Unetr: Transformers for 3d medical image
  segmentation,'' in [{\em Proceedings of the IEEE/CVF winter conference on
  applications of computer vision}{\nolinebreak\hspace{0.1em}]},   574--584
  (2022).

\bibitem{yang2022quantification}
Yang, Q., Yu, X., Lee, H.~H., Tang, Y., Bao, S., Gravenstein, K.~S., Moore,
  A.~Z., Makrogiannis, S., Ferrucci, L., and Landman, B.~A., ``Quantification
  of muscle, bones, and fat on single slice thigh ct,'' in [{\em Medical
  Imaging 2022: Image Processing}{\nolinebreak\hspace{0.1em}]},   {\bf 12032},
  422--429, SPIE (2022).

\bibitem{yu2023longitudinal}
Yu, X., Tang, Y., Yang, Q., Lee, H.~H., Gao, R., Bao, S., Moore, A.~Z.,
  Ferrucci, L., and Landman, B.~A., ``Longitudinal variability analysis on
  low-dose abdominal ct with deep learning-based segmentation,'' in [{\em
  Medical Imaging 2023: Image Processing}{\nolinebreak\hspace{0.1em}]},   {\bf
  12464},  477--483, SPIE (2023).

\bibitem{yu2020deep}
Yu, X., Lou, B., Zhang, D., Winkel, D., Arrahmane, N., Diallo, M., Meng, T.,
  von Busch, H., Grimm, R., Kiefer, B., et~al., ``Deep attentive panoptic model
  for prostate cancer detection using biparametric mri scans,'' in [{\em
  Medical Image Computing and Computer Assisted Intervention--MICCAI 2020: 23rd
  International Conference, Lima, Peru, October 4--8, 2020, Proceedings, Part
  IV 23}{\nolinebreak\hspace{0.1em}]},   594--604, Springer (2020).

\bibitem{yang2022label}
Yang, Q., Yu, X., Lee, H.~H., Tang, Y., Bao, S., Gravenstein, K.~S., Moore,
  A.~Z., Makrogiannis, S., Ferrucci, L., and Landman, B.~A., ``Label efficient
  segmentation of single slice thigh ct with two-stage pseudo labels,'' {\em
  Journal of Medical Imaging}~{\bf 9}(5),  052405--052405 (2022).

\bibitem{yu2022reducing}
Yu, X., Yang, Q., Tang, Y., Gao, R., Bao, S., Cai, L.~Y., Lee, H.~H., Huo, Y.,
  Moore, A.~Z., Ferrucci, L., et~al., ``Reducing positional variance in
  cross-sectional abdominal ct slices with deep conditional generative
  models,'' in [{\em International Conference on Medical Image Computing and
  Computer-Assisted Intervention}{\nolinebreak\hspace{0.1em}]},   202--212,
  Springer (2022).

\bibitem{yang2023single}
Yang, Q., Yu, X., Lee, H.~H., Cai, L.~Y., Xu, K., Bao, S., Huo, Y., Moore,
  A.~Z., Makrogiannis, S., Ferrucci, L., et~al., ``Single slice thigh ct muscle
  group segmentation with domain adaptation and self-training,'' {\em Journal
  of Medical Imaging}~{\bf 10}(4),  044001--044001 (2023).

\bibitem{lee2023scaling}
Lee, H.~H., Liu, Q., Bao, S., Yang, Q., Yu, X., Cai, L.~Y., Li, T.~Z., Huo, Y.,
  Koutsoukos, X., and Landman, B.~A., ``Scaling up 3d kernels with bayesian
  frequency re-parameterization for medical image segmentation,'' in [{\em
  International Conference on Medical Image Computing and Computer-Assisted
  Intervention}{\nolinebreak\hspace{0.1em}]},   632--641, Springer (2023).

\bibitem{yu2022accelerating}
Yu, X., Tang, Y., Yang, Q., Lee, H.~H., Bao, S., Moore, A.~Z., Ferrucci, L.,
  and Landman, B.~A., ``Accelerating 2d abdominal organ segmentation with
  active learning,'' in [{\em Medical Imaging 2022: Image
  Processing}{\nolinebreak\hspace{0.1em}]},   {\bf 12032},  893--899, SPIE
  (2022).

\bibitem{de2015deep}
de~Brebisson, A. and Montana, G., ``Deep neural networks for anatomical brain
  segmentation,'' in [{\em Proceedings of the IEEE conference on computer
  vision and pattern recognition workshops}{\nolinebreak\hspace{0.1em}]},
  20--28 (2015).

\bibitem{wachinger2018deepnat}
Wachinger, C., Reuter, M., and Klein, T., ``Deepnat: Deep convolutional neural
  network for segmenting neuroanatomy,'' {\em NeuroImage}~{\bf 170},  434--445
  (2018).

\bibitem{dey2018compnet}
Dey, R. and Hong, Y., ``Compnet: Complementary segmentation network for brain
  mri extraction,'' in [{\em Medical Image Computing and Computer Assisted
  Intervention--MICCAI 2018: 21st International Conference, Granada, Spain,
  September 16-20, 2018, Proceedings, Part III
  11}{\nolinebreak\hspace{0.1em}]},   628--636, Springer (2018).

\bibitem{cciccek20163d}
{\c{C}}i{\c{c}}ek, {\"O}., Abdulkadir, A., Lienkamp, S.~S., Brox, T., and
  Ronneberger, O., ``3d u-net: learning dense volumetric segmentation from
  sparse annotation,'' in [{\em Medical Image Computing and Computer-Assisted
  Intervention--MICCAI 2016: 19th International Conference, Athens, Greece,
  October 17-21, 2016, Proceedings, Part II 19}{\nolinebreak\hspace{0.1em}]},
  424--432, Springer (2016).

\bibitem{liu2021swin}
Liu, Z., Lin, Y., Cao, Y., Hu, H., Wei, Y., Zhang, Z., Lin, S., and Guo, B.,
  ``Swin transformer: Hierarchical vision transformer using shifted windows,''
  {\em Proceedings of the IEEE/CVF International Conference on Computer Vision}
   (2021).

\bibitem{dosovitskiy2020image}
Dosovitskiy, A., Beyer, L., Kolesnikov, A., Weissenborn, D., Zhai, X.,
  Unterthiner, T., Dehghani, M., Minderer, M., Heigold, G., Gelly, S., et~al.,
  ``An image is worth 16x16 words: Transformers for image recognition at
  scale,'' in [{\em International Conference on Learning
  Representations}{\nolinebreak\hspace{0.1em}]},  (2020).

\bibitem{tang2022self}
Tang, Y., Yang, D., Li, W., Roth, H.~R., Landman, B., Xu, D., Nath, V., and
  Hatamizadeh, A., ``Self-supervised pre-training of swin transformers for 3d
  medical image analysis,'' in [{\em Proceedings of the IEEE/CVF Conference on
  Computer Vision and Pattern Recognition}{\nolinebreak\hspace{0.1em}]},
  20730--20740 (2022).

\bibitem{yu2023unest}
Yu, X., Yang, Q., Zhou, Y., Cai, L.~Y., Gao, R., Lee, H.~H., Li, T., Bao, S.,
  Xu, Z., Lasko, T.~A., et~al., ``Unest: local spatial representation learning
  with hierarchical transformer for efficient medical segmentation,'' {\em
  Medical Image Analysis} ,  102939 (2023).

\bibitem{marcus2007open}
Marcus, D.~S., Wang, T.~H., Parker, J., Csernansky, J.~G., Morris, J.~C., and
  Buckner, R.~L., ``Open access series of imaging studies (oasis):
  cross-sectional mri data in young, middle aged, nondemented, and demented
  older adults,'' {\em Journal of cognitive neuroscience}  (2007).

\bibitem{asman2014hierarchical}
Asman, A.~J. and Landman, B.~A., ``Hierarchical performance estimation in the
  statistical label fusion framework,'' {\em Medical image analysis}~{\bf
  18}(7),  1070--1081 (2014).

\bibitem{evans19933d}
Evans, A.~C., Collins, D.~L., Mills, S., Brown, E.~D., Kelly, R.~L., and
  Peters, T.~M., ``3d statistical neuroanatomical models from 305 mri
  volumes,'' in [{\em 1993 IEEE conference record nuclear science symposium and
  medical imaging conference}{\nolinebreak\hspace{0.1em}]},   1813--1817, IEEE
  (1993).

\bibitem{segonne2004hybrid}
S{\'e}gonne, F., Dale, A.~M., Busa, E., Glessner, M., Salat, D., Hahn, H.~K.,
  and Fischl, B., ``A hybrid approach to the skull stripping problem in mri,''
  {\em Neuroimage}~{\bf 22}(3),  1060--1075 (2004).

\end{thebibliography}
\bibliographystyle{spiebib} 

\end{document}